\documentclass[twocolumn,prl,showpacs,amsmath,amssymb,floatfix]{revtex4}
\usepackage{graphicx}
\usepackage{hyperref} 

\usepackage{color}
\usepackage{ulem}
\usepackage{float}
\usepackage{ulem}
\newcommand{\Tr}{\mathrm{Tr}}

\newcommand{\br}{\mathbf{r}}

\newcommand{\bk}{\mathbf{k}}

\newcommand{\bn}{\begin{equation}}
\newcommand{\ee}{\end{equation}}
\newcommand{\bga}{\begin{eqnarray}}
\newcommand{\eda}{\end{eqnarray}}

\newcommand{\diff}{\text{d}}
\newcommand{\eps}{\epsilon}

\newcommand{\chalmersMC}{$^1$Department of Microtechnology and Nanoscience, MC2,
Chalmers University of Technology,
SE-41296 G\"{o}teborg, Sweden}

\begin{document}
\title{An exchange functional that tests the robustness of the
 plasmon description of the van der Waals density functional  
}
\author{Kristian Berland}
\email{berland@chalmers.se}
\author{Per Hyldgaard}
\email{hyldgaar@chalmers.se}
\affiliation{\chalmersMC}
\date{\today}
\begin{abstract}
Is the plasmon description within the non-local correlation of the van der Waals density functional by Dion and coworkers (vdW-DF1) robust enough to describe all exchange-correlation components? 
To address this question, we design an exchange functional, termed LV-PW86r based on this plasmon description as well as recent analysis on exchange in the large $s$-regime. 
In the regime
with reduced gradients $s=|\nabla n|/2n k_{\rm F}(n)$ smaller than $\approx 2.5$, dominating
the  non-local correlation part of the binding energy, the enhancement factor $F_x(s)$ 
closely resembles the Langreth-Vosko screened exchange. 
In the $s$-regime beyond, dominated by exchange, $F_x(s)$ passes smoothly over to the revised Perdew-Wang-86 form.
Our tests indicate that vdW-DF1(LV-PW86r) produces accurate separations and binding energies of the S22 data set of molecular dimers
as well as accurate lattice constants of layered materials and tightly-bound solids. 
These results suggest that vdW-DF1 has a good plasmon description in the low-to-moderate $s$-regime. 
\end{abstract}
\maketitle

van der Waals density functionals (vdW-DFs)~\cite{Rydberg:Tractable,Rydberg:Layered,Dion:vdW,vdW:selfCons,vdWDF2,Review:vdW}
include London dispersion forces within density functional theory. 
Based on first principles, they successfully describe sparse matter in its many forms, from clusters of small molecules~\cite{DimersNew} to the cohesion of layered materials~\cite{graphane,Londero1,Layered:Risto} and adsorption on surfaces~\cite{butane:Cu(111),H2onCu111,H2onCu111_B,adenineGraphene}.

The vdW-DF framework rests on the adiabatic-connection formula~\cite{rev:SpinDensity,Langreth:WaveVector} and develops the non-local correlation in terms 
of the semi-local exchange-correlation hole in a plasmon picture. 
Standard vdW-DFs builds on the generalized-gradient approximation (GGA) to account for the (semi-)local exchange-correlation energy (in an {\it outer functional}) as well for parameterizing the plasmons with an~{\it inner functional}~\cite{vdWDF2}.
While GGA succeeds in describing many kinds of dense matter, its restricted form
 brings about 
 ambiguities~\cite{PBEsol};
 making for instance some versions better for molecules and others better for solids~\cite{Burke:perspectives}.
Owing to its GGA roots, vdW-DFs inherit some of the ambiguities of GGA. 
Strikingly so for vdW-DF1 which generally has good binding energies, but whose exchange partner revPBE~\cite{GGA:PBE,GGA:revPBE,Rydberg:Layered,Dion:vdW}
 brings about a chronic overestimation of separations.
   Because of this limitation,  several alternative exchange partners for vdW-DF1 correlation have been 
developed~\cite{Cooper2009,Klimes,vdW:solids} and explored~\cite{Risto:linar,molcrys2,Londero1,C60onAu(111),Graziano:Layered}.

The inner functional of the non-local correlation of vdW-DF1 relies on the Langreth-Vosko (LV) {\it screened exchange}, stemming from a diagrammatic expansion of linear response theory~\cite{LV}.
In vdW-DF2,~\cite{vdWDF2} this functional is replaced with the gradient correction in the  large-$Z$ limit to make the functional more appropriate for atoms and molecules~\cite{Elliott_Burke:B88}.
The vdW-DF2 correlation is matched with a slightly refined version of the PW86 exchange functional (PW86r)~\cite{PW86,ExcEnergy}.
vdW-DF2 indeed performs well for systems where small molecules are involved~\cite{vdWDF2,H2onCu111,noblegas_metal,butane:Cu(111)}.
Yet vdW-DF1 has a better asymptotic behavior~\cite{Vydrov:LocPol} and tends to produce better adsorption energies for bigger molecules~\cite{Bjork_aromatic,molcrys2,berland:C60ads}. 
These properties indicate that the non-local correlation of vdW-DF1 may have a better transferability across length scales, making it a promising starting point for refining the account of sparse matter.
Especially so if emphasis is on solids~\cite{Ziambaras,Londero1,vdW:solids} and bigger molecules, because the LV exchange description is appropriate for a slowly varying electron gas. 
\begin{figure}[h!]
  \begin{center}
    \includegraphics[width=8cm]{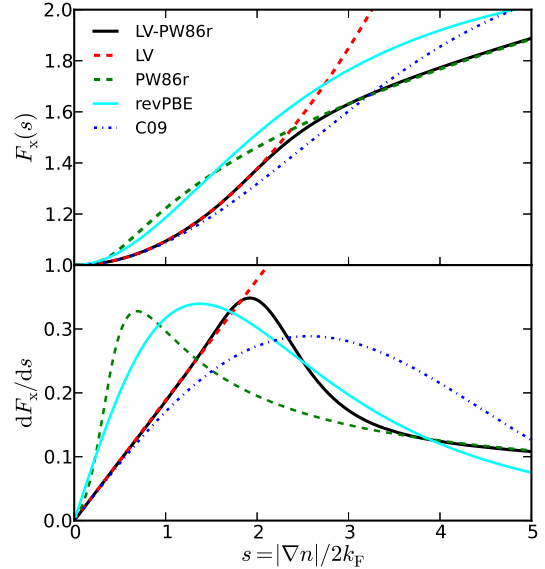}
  \end{center}
  \caption{Exchange enhancement factor $F_x(s)$ of GGA functionals. The lower panel shows $\diff F_x(s) / \diff s$ 
    which is important for binding separations. 
  This letter presents the LV-PW86r exchange functional (indicated by the thick full curve) as exchange partner for vdW-DF1 correlation.} 
  \label{fig:Fs}
 \end{figure}

This letter is motivated by the question: Is the plasmonic description that underpins the vdW-DF1 account of non-local correlation capable of 
an accurate specification of all vdW-DF functional components. 
The perspective of having such a unified plasmon representation for the functional form is that
one can fully exploit the conservation of the electron response that is built into vdW-DF~\cite{Dion:vdW}.
To address this question, we here design and test an exchange functional partner to vdW-DF1 correlation that matches the vdW-DF inner functional to the greatest extent that we deem feasible and relevant.

The total exchange-correlation energy in the vdW-DF framework is given by  
 \begin{align} 
   E^{\rm vdWDF}_{\rm xc} &=  \int_0^\infty \frac{\diff u}{2 \pi} \Tr \left[ \ln  \right(\nabla \eps \cdot  \nabla G\left)\right]  - E_{\rm self}- \delta E_{\rm xc}
  \label{eq:nl_start}\\
  \delta E_{\rm xc} = &   \int_0^\infty \frac{\diff u}{2 \pi} \Tr   \eps  -E_{\rm self}  - \underbrace{\left( E^{\rm GGA}_{\rm x} + E_{c}^{\rm LDA}  \right)}_{E^0_{\rm xc}} \,.
 \end{align} 
Here $E_{\rm self}$ is the self-energy, $E^{\rm GGA}_{\rm x}$ is the GGA exchange energy,  $E^{\rm LDA}_{\rm c}$ is the correlation energy in the local density approximation.   The term $\delta E_{\rm xc}$ 
 can compensate for 
 energetic contributions in the semi-local exchange-correlation energy that may be lost with an approximate scalar dielectric function $\epsilon$~\cite{Dion:vdW}. 
 The non-local correlation energy in vdW-DF1 and vdW-DF2 is $E^{\rm nl}_{c} = E^{\rm vdWDF}_{\rm xc} - E^0_{\rm xc}$ expanded to second order in $S=1-1/\epsilon$.
  If the inner and outer exchange functionals are the same, $\delta E_{\rm xc} =0$, the full exchange-correlation energy formally depend purely on a longitudinal projection of the dielectric function and the trivial self energy (Eq.~\ref{eq:nl_start}). The exchange-correlation hole is then automatically conserved \footnote{Ref.~\onlinecite{Dion:vdW}, describing the plasmon-pole design of vdW-DF, observes that $S=1-1/\epsilon$ 
  is finite for $\omega\neq 0$  and notes the implications for charge conservation. 
  The $\ln(\nabla\epsilon\nabla G)$ term of  Eq.~\ref{eq:nl_start} 
  involves a longitudinal projection. Conservation
  of  the corresponding exchange-correlation hole
  follows because a spatial average (of the hole)
  constitutes a zero-momentum evaluation.
  A full presentation is in preparation, details are summarized in 
  http://meetings.aps.org/Meeting/MAR13/Event/183255.
  }.

Exact matching between the inner and outer functional for all $s$ is not a good option with the vdW-DF1 description. 
The  LV form is not a good exchange in the large $s$-regime because its enhancement factor is aggressive, and, for example, noise sensitive~\cite{Londero2}.
The value of $E^{\rm nl}_{\rm c}$ is specified by 
the scaled separation $d(\br)=q_0(\br)|\br - \br'|$, where $q_0=- 4\pi\left( \varepsilon^{\rm LV}_x + \varepsilon_{\rm c}^{\rm LDA}\right)$ depends on the LDA correlation and LV exchange per particle~\cite{Dion:vdW,vdW:selfCons,vdWDF2}. 
The rapidly increasing $q_0 \propto s^2$ in the large $s$-limit implies that scaled separations diverge, tuning out $E^{\rm nl}_{\rm c}$. 
 This effect 
can be interpreted as a soft cutoff. 
We find that introducing a hard cutoff at $s=2.5$ does not greatly affect the non-local correlation energy at binding separation. 
For the outer exchange, we will thus let the enhancement factor 
roll over from the LV form to that of the PW86r form at $s\approx 2.5$.
PW86r~\cite{PW86,ExcEnergy} is chosen for the medium-to-large $s$-regime, because it was designed by imposing conservation of the exchange hole arising in the gradient expansion approximation (GEA) and 
a strong case has been made for its large $s$ form~\cite{Kannemann_Becke,ExcEnergy}. In effect, we bridge the LV form with PW86r creating LV-PW86r. 

Figure~\ref{fig:Fs} displays the enhancement factor of 
LV-PW86r exchange functional $F^{\rm LV-PW86r}(s)$ 
alongside corresponding ones for a few other proposed exchange functionals for vdW-DFs. 
This enhancement factor effectively splines the Langreth-Vosko gradient expansion~\cite{LV,Dion:vdW} $F^{\rm LV}(s) = 1 + \mu_{\rm LV}s^2$, where $\mu_{\rm LV} = -Z_{\rm ab}/9$, with the PW86r exchange enhancement factor $F^{\rm PW86r}_{\rm x}(s)  = (1 +as^2 +   b s^4 +cs^6)^{1/15}$ as follows
\begin{equation}
  F_x(s) = \left(\frac{1}{1+ \alpha s^6} \right)[1 + \mu_{\rm LV}s^2] +\left( \frac{\alpha s^6}{\beta+ \alpha s^6} \right) F^{\rm PW86r}_{\rm x}(s)\,.
  \label{eq:Fx}
 \end{equation}
The two rational functions inside the brackets tune out the LV form in favor of the PW86r form as $s$ increases. 
Since these tuning parameters are expressed in terms of sixth powers of $s$, they secure that the inner and outer exchange functional match in the low-to-moderate $s$-regime. 

The parameters $\alpha$ and $\beta$ are determined by least-squares fitting of $F_x$ to the $1 + \mu_{\rm LV}s^2$ form of LV in the  $0<s<2$ region and to that of PW86r in  the $4<s<10$ region. An equal quadratic weight to both regions results in $\alpha = 0.02178$ and $\beta =1.15$. 
The choice of these regions entails keeping the LV form up to about the point where the $F^{\rm PW86r}_{\rm x}(s)$ form crosses $F^{\rm LV}_{\rm x}(s)$.



\begin{figure}[h!]
  \begin{center}
    \includegraphics[width=8.6cm]{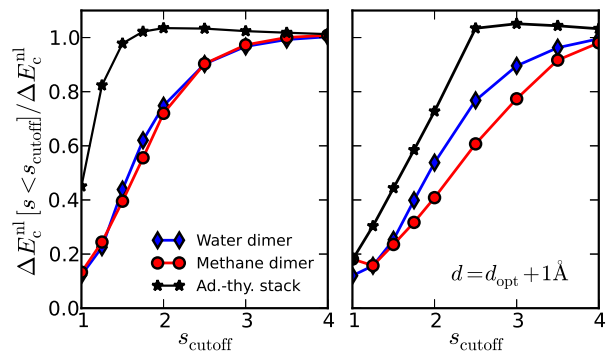}
  \end{center}
  \caption{Non-local correlation part of  interaction energy $\Delta E^{\rm nl}_{\rm c}$
  at optimal separation and 1\,\AA\, beyond for three molecular pairs of the S22 data set as a function of $s_{\rm cutoff}$: water dimer, methane dimer, and stacked adenine-thymine pair. 
  }
  \label{fig:Temp}
\end{figure}

\begin{figure*}[t!]  
    \includegraphics[width=1.0\textwidth]{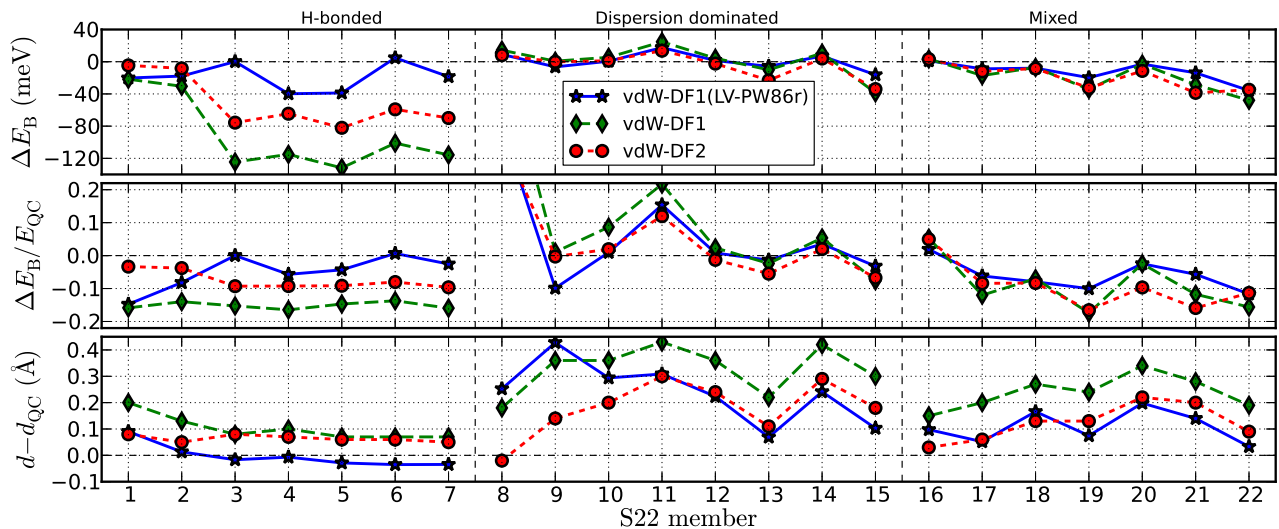}
    \caption{Results for the S22 set of molecular dimers compared with quantum chemistry (QC) results.  Enumeration as in Ref.~\onlinecite{vdWDF2}. 
Top panel compares the energetic deviation $\Delta E_B = E^{\rm vdWDF} - E_{\rm QC}$, mid panel, the relative energetic deviation, and the bottom panel, deviation from QC separation. 
  The binding energies for the methane dimer (spiking at 8) are 32, 36, 30, 23 meV for vdW-DF1(LV-PW86r), vdW-DF1, vdW-DF2, and QC. } 
  \label{fig:S22}
\end{figure*}

Figure~\ref{fig:Temp} exemplifies how low-to-moderate $s$ values dominate the non-local correlation part of the molecular interaction energy, $\Delta E^{\rm nl}_{\rm c}$ 
\footnote{This energy is the difference between the non-local correlation energy of the full system and that of the reference system(s) $\Delta E^{\rm nl}_{\rm c} =  E^{\rm nl}_{\rm c}[\text{main}] - E^{\rm nl}_{\rm c}[\text{ref}]$. Only density regions $n(\br)$ with $s<s_{\rm cutoff}$ contribute. This $s$-analysis relies on a modification of our in-house vdW-DF code detailed in Refs.~\onlinecite{molcrys1,berland:C60ads}.}.
 The left panel shows the fraction of the full $\Delta E^{\rm nl}_{\rm c}$ 
 generated
 as a function of $s_{\rm cutoff}$ for three members of the S22 data set~\cite{S22}: the water-dimer, the methane-dimer, and the stacked adenine-thymine pair at their optimal separation. 
Most of the energy is accounted for with $s_{\rm cutoff} \approx 1.5$ ($s_{\rm cutoff} \approx 2.5$),
for the stacked aromatic molecules~(for the tiny water and methane molecules)~\footnote{ Ref~\onlinecite{berland:C60ads} shows that the non-local correlation energy is very sensitive to low-density regions. 
This is consistent with the presented results 
when we consider that much of $\Delta E^{\rm}_{\rm c}$ at binding separation arise 
from density saddle-point regions between the molecules.}.
Note that, exchange may still contribute significantly to the binding energy for higher $s$ values~\cite{ExcEnergy}.
The right panel shows that if the molecules are pulled 1\,\AA\,apart (along center-of-mass line), larger $s$ values become more important. Still, density regions with $s<s_{\rm cutoff} \approx 2.5$ account for the brunt of the energy (though less so for the methane dimer). 
The left and right panels of Fig.~\ref{fig:Temp}, taken together, indicate that, with the design, good consistency is achieved at binding separation, but less so for the shape of potential energy curve of small molecules. 

Since the effective cutoff of the  non-local correlation in vdW-DF1 is soft, implicit, and system dependent, we can not identify exactly at which $s$ it is natural to tune out the LV form. 
Even so, because the LV form and PW86r cross at $s\approx2.5$, hastening or postponing this crossover could lead to a contrived enhancement factor $F_x(s)$. 
A softer crossover is an option, making $F_x(s)$ more similar to $F^{\rm C09}_x(s)$ in the $1<s<3$ region. 
However, our choice is guided by the wish to use the same plasmonic design in all functional components 
up to as large $s$ as feasible.

Figure~\ref{fig:S22} displays our benchmarking~\footnote{The planewave DFT code \textsc{Quantum Espresso} is used~\cite{QE}. For the S22 calculations, we use a planewave cutoff of 50 Ry and density cutoff of 500 Ry.  LV-PW86r calculations rely on ultrasoft pseudopotentials based on PBEsol~\cite{PBEsol},  while vdW-DF1 and vdW-DF2 rely on PBE. 
Soler's algorithm is used to evaluate the non-local correlation~\cite{fast:Soler}.}  based on the S22 data set of molecular dimers~\cite{S22} for the proposed functional combination vdW-DF1(LV-PW86r) compared with results of vdW-DF1 and vdW-DF2. Following Ref.~\onlinecite{vdWDF2}, the quantum chemistry (QC) result are based on the CSSD(T) calculations of Ref.~\onlinecite{S22:5}. 

On average, vdW-DF1(LV-PW86r) performs marginally better than vdW-DF2 does.  
For the binding energy, we find a mean absolute relative deviation of 7\% for vdW-DF1(LV-PW86r) compared to 9\% for vdW-DF2\footnote{Ref.~\onlinecite{vdWDF2} reports mean absolute relative deviation of 8\% for vdW-DF2.} and 13\% for vdW-DF1. 
For the separations, the mean absolute deviations are respectively 0.13\,\AA, 0.14\,\AA, and 0.23\,\AA. 
Though the overall performance of vdW-DF2 and vdW-DF1(LV-PW86r) are similar, their trends differ. 
vdW-DF2 prevails for small molecules and vdW-DF1(LV-PW86r) for the bigger aromatic ones. 

The significant overestimation of separations for the methane and ethene dimer (8 and 9)  when using  vdW-DF1(LV-PW86r) we interpret as a consequence of keeping the LV form as long as feasible. LV-PW86r has the largest $\diff F/\diff s$ in the $s\approx 2$ region as revealed by the lower panel of Fig.~\ref{fig:Fs}. Larger $s$ values matter more for small pointy molecules as Fig.~\ref{fig:Temp} exemplifies. 
Moreover, the agreement between the outer and the inner functional is necessarily poorer for these small molecules. 

\begin{table}[t!]
  \caption{Lattice constants ({\text \AA}) calculated with different versions and variants of vdW-DF.}
  \begin{ruledtabular}
\begin{tabular}{llllllll}
  \vspace{0.1cm} xc &vdW-DF1    & vdW-DF2  &  C09\hspace{0.3cm}  &  LV-PW86r\hspace{0.3cm} &  Exp.\footnotemark[1]  \\
Li   &  3.47  & 3.38 &  3.44 & 3.48  &   3.45    \\
Na   &  4.20  & 4.14 &  4.22 & 4.24  &   4.21    \\
Al & 4.08  & 4.09  & 4.02  & 4.02  &   4.02      \\
  Cu &   3.70  &  3.74  & 3.58   &3.57  &   3.57  \\
Ag & 4.24  & 4.31  & 4.05   & 4.07   &  4.06  \\
Au & 4.26 & 4.36 & 4.10 & 4.10 & 4.06 \\
C & 3.59  & 3.62 & 3.56  &  3.56   &  3.54        \\
Si & 5.50  & 5.54   &  5.43 &  5.43   &  5.42      \\
Pb &  5.15    &  5.24  &  4.92   &    4.94    & 4.91      \\
GaAs & 5.83 &  5.92   & 5.64     &   5.67    &  5.64 \\
InAs  & 6.19 &  6.38  &  6.09  & 6.11   &  6.04    \\
Pd      &   4.01     & 4.09   &   3.88      &   3.89       &  3.88 \\
MAD &0.12  &  0.19 & 0.014  & 0.023 & -  \\
\end{tabular}
\footnotemark[1]{Experimental data as listed in Ref.~\onlinecite{lattice_constants}}
\label{tab:solids}
\end{ruledtabular}
\end{table} 
\begin{table}
  \caption{Lattice constants ({\text \AA}) and interlayer binding energy per area (meV/$\text{\AA}^2$) calculated with different versions and variants of vdW-DF compared with experimental data.}
  \begin{ruledtabular}
\begin{tabular}{llllllll}
  \vspace{0.1cm} xc & vdW-DF1    & vdW-DF2 &  C09\hspace{0.3cm}  &  
LV-PW86r\hspace{0.3cm} &  Exp \\
Graphite \\
\hspace{0.2cm}  $a$ & 2.47   & 2.47   & 2.46  & 2.46    &   2.46\footnotemark[1]    \\
\hspace{0.2cm}  $c$ & 7.13   & 7.02    & 6.44    & 6.53    &  6.68\footnotemark[1]  \\
\hspace{0.2cm}  $E/A$ & 20.6  & 20.0   &  29.3  & 25.1 & 17\footnotemark[2]23\footnotemark[3]
 \\
BN  \\
\hspace{0.2cm}  $a$ & 2.51  & 2.53  &     2.51  & 2.51   &  2.51\footnotemark[4]      \\
\hspace{0.2cm}  $c$ &  7.00   &  6.92 & 6.31   & 6.39   &  6.60\footnotemark[4]       \\
\hspace{0.2cm}  $E_{\rm }/A$ & 20.0 & 18.5 &  28.5 &  24.1  &  -  \\
MoS2\\
\hspace{0.2cm}  $a$    & 3.24   & 3.29  &   3.15    & 3.15    &   3.16\footnotemark[5]    \\
\hspace{0.2cm}  $c$ 	& 13.03  & 12.83  &  12.13   & 12.25    &  12.29\footnotemark[5]  \\
\hspace{0.2cm}  $E_{\rm }/A$  &  18.7   & 19.3 &  29.4   & 24.6   & -  \\
WSe2\\
\hspace{0.2cm}  $a$ & 3.38     & 3.45  & 3.28    &  3.29    &  3.28\footnotemark[6]      \\
\hspace{0.2cm}  $c$ &  13.90   & 13.74  & 12.91   & 13.01   & 12.96\footnotemark[6]      \\
\hspace{0.2cm}  $E_{\rm }/A$ &  17.8  & 18.0  &   29.0      &   24.5      &   -  \\
$\alpha$-PbO\\
\hspace{0.2cm}  $a$ &        4.14  & 4.18  & 4.01    &      4.02  &   3.96\footnotemark[7]       \\
\hspace{0.2cm}  $c$ &        5.61  &  5.46  & 4.83   &      4.92  &   5.01\footnotemark[7]       \\
\hspace{0.2cm}  $E_{\rm }/A$ & 13.3  & 13.5    &   26.1   &    21.2  & - 
\end{tabular}
\footnotemark[1]{Ref.~\onlinecite{Baskin:graphite}}
\footnotemark[2]{Ref.~\onlinecite{exp_graphite:Liu}}
\footnotemark[3]{Ref.~\onlinecite{exp_graphite:Hertel} (low $T$)}
\footnotemark[4]{Ref.~\onlinecite{Paszkowicz:BN} (low $T$)}
\footnotemark[5]{Ref.~\onlinecite{Bronsema:MoS2}}
\footnotemark[6]{Ref.~\onlinecite{Schutte:WSe2}}
\footnotemark[7]{Ref.~\onlinecite{Leciejewicz:PbO}}

\label{tab:layered}
\end{ruledtabular}
\end{table}  

For the opposite regime, that of tightly-bound solids, \footnote{A planewave cutoff of 50 Ryd. density cutoff of 500 Ryd. and $\bk$-points sampling of $16\times16\times16$ is used.} 
table~\ref{tab:solids} confirms our expectations of good lattice constants of vdW-DF1(LV-PW86r)~\footnote{
Here we determine some atomization energies as in Ref.~\onlinecite{Ziambaras}, asserting the energy for spin polarization at the GGA level. This procedure indicates that vdW-DF1(LV-PW86r) improves the atomization energies over vdW-DF1.   Examples: Li,  vdW-DF1: 1.41 eV, vdW-DF1(PW86r): 1.47 eV, Exp: {\bf 1.67} eV;   Al:  2.95, 3.56, {\bf 3.44}    Ag: 2.10,   2.96, {\bf 2.97}.}.
Klime\v{s} 
 and coworkers~\cite{vdW:solids} investigated the effect of using different vdW-DFs, 
including variants based on semi-empirical fitting~\cite{Klimes}.
Their benchmarking revealed that vdW-DF1 and vdW-DF2 typically overestimate lattice constants, especially for heavier elements towards the right end of the periodic system. 
Omitted in their study, we also present results for vdW-DF1(C09), where C09 is the exchange functional designed by Cooper~\cite{Cooper2009} for vdW-DF1. 

As in the work of Cooper~\cite{Cooper2009}, Klime\v{s}~et.~al.~\cite{vdW:solids} was inspired by the design logic of PBEsol~\cite{PBEsol} to construct an exchange functional named optB86b as exchange partner for vdW-DF1. 
The logic in C09 and optB86b is to follow the second order small-$s$ expansion factor 
$F(s) = 1 + \mu s^2 + \cdots$ in a wide region of $s$ values. The lowering of 
$\diff F(s)/\diff s$, relative to PW86r and revPBE, improves lattice constants significantly. 
LV-PW86r benefits from the same mechanisms as a byproduct of stressing consistency in the plasmon description. 






Five layered systems, selected for variety and technological interest, provide our final testing ground. 
Table~\ref{tab:layered} presents our results 
and available experimental data 
\footnote{vdW-DF1 is known to underestimate the elastic coefficient $C_{33}$ of graphite\cite{Ziambaras}: 27~GPa compared to experimental of 37-39~GPa~\cite{Theory:Elastic_graphene}. vdW-DF1(LV-PW86r) predicts 40~GPa.}.
Experimental numbers for interlayer binding energies are scarce, but it has been suggested to compare them to random-phase approximation (RPA)~\cite{Layered:Risto,vdW:ready} calculations.
For the in-plane lattice constant $a$, trends are similar to those of the regular solids. 
In the out-of-plane direction, vdW-DF1 significantly overestimates the lattice constant $c$ as expected. vdW-DF2 improves interlayer separations somewhat, but less than one might expect from the S22 results. 
Since vdW-DF1 has a stronger non-local correlation effect, we attribute most of this lacking improvement to the fact that the geometry of layered materials is different than for dimers, noting that the distributions of electrons are of central importance~\cite{berland:C60ads}.
vdW-DF1(C09) performs quite well, but underestimates $c$. 
vdW-DF1(LV-PW86r) provides the best results. 


The results presented here indicate that vdW-DF1(LV-PW86r) performs well for solids, layered materials, and aromatic molecules. 
This is in line with its emphasis on good properties of a slowly varying electron gas. In contrast, vdW-DF2 puts emphasis on small molecules. 
vdW-DF1(LV-PW86r) is constructed to test the ability of the vdW-DF1 plasmon description to specify all full functional components. The promising results are encouraging for developing 
van der Waals functionals following a strategy demanding a unified plasmon description. 

We thank E.~Schr\"oder, B.~I.~Lundqvist and K.~Lee for valuable input. 
The Swedish Research Council (VR) and the Chalmers Area of Advance supported this work. 
The Swedish National Infrastructure for Computing (SNIC) at the C3SE and HPC2N provided computer time.

\bibliographystyle{apsrev4-1}
\bibliography{bib2}

\end{document}